\documentstyle[11pt]{article}

\begin{document}

\title{Phase Transitions and Adsorption Isotherm in Multilayer
Adsorbates with Lateral Interactions}

\author{Sonia B. Casal$^1$\thanks{Electronic mail:
scasal@mdp.edu.ar}, Horacio S. Wio$^2$\thanks{Electronic mail:
wio@cab.cnea.gov.ar} and S. Mangioni$^1$\thanks{Electronic mail:
smangio@mdp.edu.ar}\\
(1) Departamento de F\'{\i}sica, Facultad de Ciencias Exactas y
Naturales, \\ Universidad Nacional de Mar del Plata,
De\'an Funes 3350, 7600 Mar del Plata, Argentina. \\
(2) Grupo de F\'{\i}sica Estad\'{\i}stica
\thanks{http://www.cab.cnea.gov.ar/Cab/invbasica/FisEstad/estadis.htm},
Centro At\'omico Bariloche (CNEA) \\ and Instituto Balseiro (CNEA
and UNC), 8400 San Carlos de Bariloche, Argentina.}


\maketitle

\begin{abstract}
We analyze here a model for an adsorbate system composed of many
layers by extending a theoretical approach used to describe
pattern formation on a monolayer of adsorbates with lateral
interactions. The approach shows, in addition to a first order
phase transition in the first layer, a transition in the second
layer together with evidence of a ``cascade" of transitions if
more layers are included. The transition profiles, showing a
staircase structure, corroborate this picture. The adsorption
isotherm that came out of this approach is in qualitative
agreement with numerical and experimental results.
\end{abstract}


\newpage

\section{Introduction}

The problem of phase transitions and ordering phenomena in
adsorbed films on crystal surfaces has attracted considerable
interest for many years (see \cite{rx} for a recent review).
Formation of ordered structures of particles adsorbed on surfaces
due to mutual (lateral) interactions has been observed in many
experiments, originating different types of adsorption isotherm
\cite{r1,r2}. Strong attractive (lateral) interactions lead to
phase separation with different coverages, characteristic of
first order phase transitions \cite{r1,r2}.

It is well known experimentally that adsorption on surfaces is not
only restricted to the formation of monolayers, but rather a
second layer can condensate on the first one, a third on the
second, and so on \cite{ry}. Numerical simulations also show
similar results \cite{rz}. The old BET theory \cite{bet}
considered the possibility of multilayer adsorption but, as it
does not include interactions among particles in the same layer,
does not yield either a phase transition or ordered structures.
So far we are not aware of theoretical models to describe this
situation. Other theoretical forms for isotherm intended to
describe such a multilayer adsorption phenomenon have been
discussed in the literature \cite{ISO}. However, such models do
not seem to be able to describe phase transitions or pattern
formation.

The aim of the present work is to introduce a simple model that
can describe multilayer adsorption qualitatively and the
formation of ordered structures. To reach this goal we exploit a
model introduced in Ref.\cite{r1} to describe a system of a
monolayer of adsorbed particles with lateral attractive
interactions leading to the formation of interfaces separating
phases with different local coverage. The new aspect is that we
extend such a model and go beyond the one layer system
considering a multilayer system (however, focusing on the case of
only two layers) including also an exchange interaction between
different layers. It is worth remarking here that such a
``macroscopic" model can be derived from a microscopic one
following a procedure similar to that used in \cite{micros}.

In what follows we present the model, show the form of the
isotherm, and discuss briefly some aspects of the critical point and
the form of the density profiles of the possible fronts. Finally
we draw some conclusions.

\section{Adsorptive multilayer system}

We adopt here a phenomenological point of view. However, a formal
derivation of the macroscopic master equation leading us to the
phenomenological equations we present here, is included in the
appendix, where we start from a microscopic description closely
following the procedure of Ref.\cite{micros}).

We adopt a continuous description for the surface, and
characterize the adsorptive species through evolution equations
for $c_{i}(x,t)$, the local coverage in the $i$-th layer (see the
appendix for the precise definition of these local coverage). The
adsorptive contribution is characterized by $k_{a}$ (a constant),
and the adsorption is only possible in the $(1-c_{i})$ free sites
of the corresponding layer. Hence, the adsorption rate is $k_{a} p
(1-c_{i})$, where $p$ is the partial pressure of the gaseous
phase. Additionally, for adsorption on the $i$-th  layer, it is
required to have free sites on the $i+1$-th layer.

The desorption process has a rate $k_{d}$, that includes
$k_{d,0}$, the desorption for noninteracting particles, and a
correction due to the lateral interactions. The strong local bond
induced by the interaction $U_1(x)$ ($U_2(x)$ for the second
layer, etc), corrects the desorption rate as: $k_{d}= k_{d,0}
\exp[U_i(x)/kT]$, where $k$ is the Boltzmann constant and $T$ the
temperature. According to the form we use to introduce such an
interaction, we are assuming that it is a {\it substratum
mediated} interaction. A desorption process in the first layer
can happen only if the corresponding site on the second layer is
free.

The potential  $U_i(x)$ produces a force $F_i= -\frac{\partial
U_i(x)}{\partial x}$, that affects the adsorbed particles inducing
a velocity $v_i=b F_i$, where $b$ is the mobility (given by
Einstein's relation $b=\frac{D}{kT}$, with $D$ the diffusion
coefficient in the $i$-th layer). The associated particle flux is
$j_i \sim c_i\,v_i$. Because the flux is only possible at the
$(1-c_i)$ free sites, its final form is $$j_i= - \frac{D}{KT}\,
c_i\, (1-c_i) \,\frac{\partial U_i(x)}{\partial x},$$ while the
diffusion flux is given by $$j_1^{dif}=-D\, \frac{\partial
c_1}{\partial x},\,\,\,\,\,\,\,\,\,\,j_2^{dif}=-D\, \frac{\partial
c_2}{\partial x},$$ for the first and the second layer
respectively.

The possible transference interaction between different layers is
characterized by a coupling parameter $k_{T}$. It involves free
sites in the final layer and occupied ones in the original layer.
Hence, the form of this contribution is $k_{T} c_{i}(1-c_{i+1})$
for transfer from a given layer to the upper one, and similarly in
other cases. In Fig. 1 we sketch the different processes entering
into our model.

Considering all the above, the evolution equations for the
different coverage can be written as
\begin{eqnarray}
\label{1} \frac{\partial }{\partial t}c_1(x,t)= k_a p
(1-c_1)(1-c_2) &-& k_{d,0}(1-c_2) c_1 \exp[U_1(x)/kT] + c_1(1-c_2)k_T  \nonumber \\
&+ & \frac{\partial }{\partial x}[\frac{D}{kT}\frac{\partial
U_1(x)}{\partial x}
c_1(1-c_1)+D\frac{\partial c_1}{\partial x}] \\
\frac{\partial }{\partial t}c_2(x,t)=k'_ap\,c_1(1-c_2) &-&
k'_{d,0}c_2 \exp[U_2(x)/kT] + k'_T c_2 (1-c_1) \nonumber \\
&+&\frac{\partial }{\partial x} [\frac{D}{kT}\frac{\partial
U_2(x)}{\partial x} c_2(1-c_2)+D\frac{\partial c_2}{\partial x}].
\end{eqnarray}

According to \cite{r1}, we can employ the following simple
approach. In thermal equilibrium the coverage $c_i(x)$ of the
i-th layer is stationary, that is $\frac{\partial c_i}{\partial
t}=0$.  The flux induced by the gradient of the potential is
balanced by the diffusion one yielding
\begin{eqnarray}
k_ap\,(1-c_1)(1-c_2)&-&k_{d,0}(1-c_2)c_1 e^{U_1(x)/kT}+
c_1(1-c_2)k_T=0 \\ k'_ap\,c_1(1-c_2)&-&k'_{d,0}c_2 e^{U_2(x)/kT} +
k'_Tc_2(1-c_1)=0.
\end{eqnarray}

If we consider the case of constant $U_i(x)$ as well as no
transfer between the layers, we have
\begin{eqnarray}
k_ap\,(1-c_1)(1-c_2)&-&k_{d,0}(1-c_2)c_1 e^{U_1/kT} = 0 \\
k'_ap\,c_1(1-c_2)&-&k'_{d,0}c_2  e^{U_2/kT} = 0.
\end{eqnarray}
We can get $c_1$ from the first expression obtaining a result
that agrees with the well known form of the Langmuir isotherm for
adsorption in a monolayer \cite{r1}
\begin{equation}
\label{piru1} c_1= \frac{1}{1+ (\frac{k_{d,0}}{k_ap}) e^{U_1/kT}}.
\label{}
\end{equation}
We can also we obtain $c_2$ from the previous system and consider
the total coverage, that can be eventually compared with the
results of the usual BET isotherm \cite{bet} or similar ones
\cite{ISO}. However, our main interest is to obtain relations and
conditions yielding the phase transition, as well as the form of
evolution fronts.

For the functional form of $U_i(\vec{x})$ we assume an attractive
(and as indicated earlier, {\it substratum mediated}) potential
among particles separated a distance $r$, that we denote by
$u_i(r)$. The potential acting on a particle located at $r$ in
the first layer is
\begin{equation}
U_1(r)= - \int u_1(r-r')c(r')\,dr'.
\end{equation}
The integration domain is the whole surface. As a first step in
our analysis we assume that we have the same potential $u(r-r')$
among particles in all layers. The function $u(r)$ depends on the
nature of the system. If the interaction radius is small compared
with the diffusion length, and the coverage is not much affected
by variations in this radius, we can approximate
\begin{equation}
\label{taylor} \int u(r-r') c_{1}(r') dr'\simeq \int
u(r-r')[c_{1}(r)+(r-r')\frac{\partial c_{1}}{\partial
r}+\frac{1}{2}(r-r')^{2}\frac{\partial^{2}c_{1}(r)} {\partial
r^{2}}+ \cdots] dr'.
\end{equation}
The spatial derivatives are evaluated at $r$. Hence we obtain
\begin{equation}\label{apro}
U_1(r)=- \int u(r-r') c_{1}(r') dr'\simeq - u_{0} c_{1} - \chi _1
\frac{\partial^{2}c_{1}}{\partial r^{2}},
\end{equation}
where the coefficients are given by
\begin{eqnarray}
u_{0}=\int u(r) dr,
\end{eqnarray}
and
\begin{eqnarray}
\chi _1=1/2 \int r^{2}u(r) dr,
\end{eqnarray}
and from symmetry considerations we have $\int r u(r) dr=0$.

In a similar way (and remembering that we assume the lateral
interaction is substratum mediated) for the second layer we get
\begin{equation}
U_2(r)=c_{1}[-\int u(r-r') c_{2}(r') dr'].
\end{equation}
After replacing $c_{2}(r')$ by its approximation (that is a
Taylor expansion around $r$ as in Eq. (\ref{taylor})), we find
\begin{equation}\label{U'}
U_2(r)=- c_{1}\int u(r-r') c_{2}(r') dr' \simeq -c_{1}[ u_{0}
c_{2}+\chi _2 \frac{\partial^{2}c_{2}}{\partial r^{2}}],
\end{equation}
where $u_{0}$ and $\chi _2$ have the same form as before.

Replacing $U_1(r)$ y $U_2(r)$ into the equations of motion
Eqs.(1,2), we get
\begin{eqnarray}
\frac{\partial }{\partial t}c_{1}(r,t)= k_{a} p (1-c_1)(1-c_2)
&-&k_{d,0}(1-c_{2}) c_{1} e^{-u_{0}c_1/kT}+ c_1(1-c_2)k_T
\nonumber
\\ &+& \frac{\partial }{\partial r} [D(1- \frac{u_{0}}{kT}
(1-c_1) c_1) \frac{\partial c_1} {\partial r}],
\end{eqnarray}
\begin{eqnarray}
\frac{\partial }{\partial t} c_{2}(r,t)= k'_{a} p c_1(1-c_2) &-&
k'_{d,0}c_{2} e^{-c_1 u_{0} c_{2}/kT}+k'_T c_2(1-c_1) \nonumber
\\ &+& \frac{\partial }{\partial r} [D(1-\frac{u_{0}}{kT}
c_{1}(1-c_{2})c_{2}) \frac{\partial c_{2}}{\partial r}].
\end{eqnarray}
In both equations we have neglected the contribution coming from
the terms $\chi _i \partial_{r}^{2}c_{i}$, as we follow Ref.
\cite{micros}, and assume they are only small contributions.

To simplify the notation we assume $D_1=D_2=D$, and we scale the
variables as follows: $\xi=x/L_{dif}$, where the diffusion length
is $L_{dif}=(D/k_{d,0})^{1/2}$; $\tau =t/t_d$, with
$t_d=1/k_{d,0}$; $\varepsilon =u_0/kT$. Finally, $\alpha
=k_ap/k_{d,0}$ characterizes the coverage in equilibrium when
$U(x)$ becomes zero. Also, $K=k_T/k_{d,0}$ is the exchange
parameter.

With the indicated scaling the evolution equations for
$c_i(\xi,\tau )$ reduce to
\begin{eqnarray}
\label{sol1c}
\frac{\partial c_1}{\partial \tau}&=&
\alpha_1(1-c_1)(1-c_2)-(1-c_2)c_1 \exp[-\varepsilon c_1]
\nonumber \\
& &- \frac{\partial }{\partial \xi}[\varepsilon c_1(1-c_1)
\frac{\partial c_1}{\partial \xi }]+\frac{\partial^2
c_1}{\partial \xi^2}+c_1(1-c_2)K
\end{eqnarray}
\begin{eqnarray}
\label{sol2c} \frac{\partial c_2}{\partial
\tau}&=&\alpha_2c_1(1-c_2)-c_2\exp[-\varepsilon c_1 c_2] \nonumber
\\
& &-\frac{\partial }{\partial \xi}[\varepsilon c_1(1-c_2)c_2
\frac{\partial c_2}{\partial \xi}]+\frac{\partial^2 c_2}{\partial
\xi ^2}+c_2(1-c_1)K.
\end{eqnarray}

As in \cite{r1}, we may consider for the stationary regimen or
thermodynamic equilibrium (and initially assuming $K=0$) that
\begin{eqnarray}
\alpha_1(1-c_1)-c_1\exp[-\varepsilon c_1]&=&0 \\
\alpha_2c_1(1-c_2)-c_2\exp[-\varepsilon c_1 c_2]&=&0.
\end{eqnarray}

The dependence of the coverages $c_i$ on $\varepsilon$,
$\alpha_1$ and $\alpha_2$ is apparent. The result is analogous to
that shown in Fig. 2 of Ref.\cite{r1}. It shows that in the plane
$(\varepsilon$,$\alpha_1)$ there are three regions. Two of them
correspond to real solutions, with homogeneous coverage, while
the intermediate region is bistable and presents a coexistence
phase with intermediate coverages. Those regions coalesce in a
critical point indicating a second order phase transition.

In equilibrium and from the diffusive term we get equations for
$c_1$ and  $c_2$:
$$1-\varepsilon c_1(1-c_1)=0,$$
$$1-\varepsilon c_1(1-c_2)c_2=0.$$
The last expressions indicate that in order to have a high
coverage in the first layer we need  $\varepsilon >4$, while for
the second layer we find a relation with the coverage of the first
one through $\varepsilon > 4/c_1$. This indicates that  $c_2$
becomes relevant only after some threshold value of the coverage
has been reached in the first layer.

The kind analysis we have just outlined, has been strongly
criticized in \cite{zhd} on the basis that the Maxwell
construction in \cite{r1} is done assuming that coverages near
the interface correspond to the spinodal points. This was
discussed and corrected in \cite{ME} using the true equilibrium
condition, that is that the chemical potentials of both phases
should be equal at the transition point. In the next Section we
use such an approach in order to determine the correct phase
diagram.

\section{Local Chemical Potential}

As indicated in the last part of the previous section and in
order to analyze the phase transition we follow \cite{ME}
defining a local chemical potential for each layer according to
\begin{equation} \label{pot1}
\varphi=-\int u(r-r')\, c_{1}(r')\, dr' + kT\,
\ln[\frac{c_{1}(r)}{1-c_{1}(r)}],
\end{equation}
\begin{equation}\label{pot2}
\varphi'=- c_{1}\,\int u(r-r') c_{2}(r') dr' + kT \,
\ln[\frac{c_{2}(r)}{1-c_{2}(r)}].
\end{equation}

Using these definitions, the equations of motion can be written as
\begin{equation}
\frac{\partial{ c_{1}}}{\partial{t}}=k_{a}p(1-c_{1}) [1-
e^{\frac{\varphi-\varphi_{0}}{kT}}]+
\frac{\partial{}}{\partial{r}}[\frac{D}{kT}
c_{1}(1-c_{1})\frac{\partial{\varphi}}{\partial{r}}],
\end{equation}
\begin{equation}
\frac{\partial{ c_{2}}}{\partial{t}}=k_{a}p c_{1}(1-c_{2})[1-
e^{\frac{\varphi'-\varphi'_{0}}{kT}}]+
\frac{\partial{}}{\partial{r}}[\frac{D}{kT}
c_{2}(1-c_{2})\frac{\partial{\varphi'}}{\partial{r}}],
\end{equation}
where
\begin{equation}
\varphi_{0}=kT \,\ln[\frac{k_{a}p}{k_{d,0}}],
\end{equation}
and
\begin{equation}
\varphi'_{0}=kT \,\ln[\frac{k'_{a}p}{k'_{d,0}}c_{1}].
\end{equation}
In equilibrium these chemical potentials are constant on all
surface regions corresponding to each layer and equal to
$\varphi_{0}$, for the first one and $\varphi'_{0}$, for the
second.

For a uniform distribution, according to Eqs.(\ref{pot1}) and
(\ref{pot2}), the chemical potentials are given by
\begin{equation}\label{pote1}
\varphi(c_{1})= - u_{0}c_{1} + kT\, \ln[\frac{c_{1}}{1-c_{1}}],
\end{equation}
and
\begin{equation}\label{pote2}
\varphi'(c_{2})= - u_{0}c_{1}c_{2} + kT\,
\ln[\frac{c_{2}}{1-c_{2}}].
\end{equation}

In figure 2 we depict the phase diagram that clearly shows three
regions: a low and a high coverage region, and a central region
corresponding to coexistence. This last region is bistable in
both layers and fulfills the conditions
$\frac{d\varphi}{dc_{1}}=0$; $\frac{d\varphi'}{dc_{2}}=0$; that
explicitly indicate
\begin{equation}
\frac{d\varphi}{dc_{1}} = - u_{0} + \frac{kT}{c_{1}(1-c_{1})} = 0,
\end{equation}
and
\begin{equation}
\frac{d\varphi'}{dc_{2}} = - u_{0}c_{1} +
\frac{kT}{c_{2}(1-c_{2})} = 0.
\end{equation}

In order that the interface separating both phases be stationary,
the Maxwell condition should be satisfied in both cases. Such a
condition can be derived for $\varphi=\varphi_{0}$ and
$\varphi'=\varphi'_{0}$, together with the approximation
indicated in Eq. (\ref{apro}). This yields the following equations
\begin{equation}
- u_{0}c_{1} + kT\, \ln(\frac{c_{1}}{1-c_{1}}) - \chi _1
\frac{\partial^{2}c_{1}}{\partial r^{2}}=\varphi_{0},
\end{equation}
and
\begin{equation}
- u_{0}c_{2} + kT \,\ln(\frac{c_{2}}{1-c_{2}}) - \chi _2
\frac{\partial^{2}c_{2}}{\partial r^{2}}=\varphi'_{0}.
\end{equation}
>From these expressions we can obtain the Maxwell condition
corresponding to the coexistence of both phases. For the first
layer we have
\begin{equation} \label{integra1}
\int_{c_{11}}^{c_{12}} (\varphi(c_{1})-\varphi_{0})\,dc_{1} = 0,
\end{equation}
where $\varphi(c_{1})$ is given by Eq. (\ref{pote1}), and
$c_{11}$, $c_{12}$ are the coverage values in the equilibrium
phases on the first layer. For the second layer we have
\begin{equation}\label{integra2}
\int_{c_{21}}^{c_{22}} (\varphi'(c_{2})-\varphi'_{0})\,dc_{2} = 0,
\end{equation}
where $\varphi(c_{2})$ is given by Eq. (\ref{pote2}), and
$c_{21}$, $c_{22}$ are the coverage values in the equilibrium
phases.

\section{Stationary coexisting states}

\subsection{First layer}

As seen above, the coexistence line between the dense and diluted
phases on the first layer is given by the Maxwell condition in
Eq. (\ref{integra1}), where
\begin{equation}
\varphi (c_1)= - u_{0} c_{1} + kT \,\ln(\frac{c_{1}}{1-c_{1}}),
\end{equation}
\begin{equation}
\varphi_{0} = kT \,\ln\frac{k_{a}p}{k_{d,0}}
\end{equation}
After integrating we find
\begin{eqnarray}
\frac{-u_{0}c_{11}^{2}}{2} &+& kT \,\ln(1-c_{11}) + kT \,c_{11}
\,\ln(\frac{c_{11}}{1-c_{11}}) - \varphi_{0} c_{11} = \nonumber \\
&=& \frac{-u_{0}c_{12}^{2}}{2} + kT \,\ln(1-c_{12}) + kT \,c_{12}
\,\ln(\frac{c_{12}}{1-c_{12}}) - \varphi_{0} c_{12}. \label{b1}
\end{eqnarray}
Here, $c_{11}$ y $c_{12}$ are the equilibrium coverages of both
phases on the first layer. They satisfy the equations
\begin{equation}\label{b2}
- u_{0}c_{11} + kT \,\ln(\frac{c_{11}}{1-c_{11}}) = \varphi_{0},
\end{equation}
\begin{equation}\label{b3}
- u_{0}c_{12} + kT \,\ln(\frac{c_{12}}{1-c_{12}}) = \varphi_{0},
\end{equation}
with $\varphi_{0}$ given by
\begin{equation}\label{b4}
\varphi_{0}= kT \,\ln(\frac{k_{a}p}{k_{d,0}}).
\end{equation}

After some algebra, where the condition $c_{11}+c_{12}=1$ arises,
we get
\begin{equation}\label{b10}
-u_{0}c_{11} + kT \,\ln(\frac{c_{11}}{1-c_{11}}) = - \frac{1}{2}
u_{0}.
\end{equation}
When comparing Eqs. (\ref{b2}) and (\ref{b10}), we see that
\begin{equation}
\varphi_{0}= - \frac{1}{2} u_{0}.
\end{equation}
Replacing  $\varphi_{0}$ by Eq. (\ref{b4}) we finally obtain
the coexistence condition on the first layer
\begin{equation}\label{isot1}
\frac{k_{a}p}{k_{d,0}}= e^{-\frac{u_{0}}{2kT}}.
\end{equation}

\subsection{Second layer}

In a similarly way for the second layer, we consider Eq.
(\ref{integra2}), where
\begin{equation}
\varphi' = - u_{0}c_{1} c_{2} + kT \,\ln
\left(\frac{c_{2}}{1-c_{2}}\right),
\end{equation}
and
\begin{equation}
\varphi'_{0} = kT \,\ln \left(\frac{k'_{a}p}{k'_{d,0}}c_{1}
\right).
\end{equation}
After integrating we get
\begin{eqnarray}
-\frac{u_{0}c_{1}c_{21}^{2}}{2} &+& kT \,\ln(1-c_{21}) + kT
\,c_{21} \,
\ln(\frac{c_{21}}{1-c_{21}}) - \varphi'_{0} c_{21}= \nonumber  \\
&=& -\frac{u_{0}c_{1}c_{22}^{2}}{2} + kT \,\ln(1-c_{22}) + kT \,
c_{22}\,\ln(\frac{c_{22}}{1-c_{22}}) - \varphi'_{0} c_{22}.
\label{b12}
\end{eqnarray}
Here $c_{21}$ and $c_{22}$ are the equilibrium coverages in both
phases of the second layer. They fulfill the equations
\begin{equation}\label{b22}
- u_{0}c_{1}c_{21} + kT \,\ln(\frac{c_{21}}{1-c_{21}}) =
\varphi'_{0},
\end{equation}
and
\begin{equation}\label{b32}
- u_{0}c_{1}c_{22} + kT \,\ln(\frac{c_{22}}{1-c_{22}}) =
\varphi'_{0},
\end{equation}
and $\varphi'_{0}$ is given by
\begin{equation}\label{b42}
\varphi'_{0}= kT \,\ln(\frac{k'_{a}p}{k'_{d,0}}c_{1}).
\end{equation}

As in the previous subsection, after some algebra where the
condition $c_{21}+c_{22}=1$ arises, we get
\begin{equation}\label{b102}
-u_{0}c_{1}c_{21} + kT \,\ln(\frac{c_{21}}{1-c_{21}}) = -
\frac{1}{2} u_{0}c_{1}.
\end{equation}
Comparing Eqs. (\ref{b22}) and (\ref{b102}), we see that
\begin{equation}
\varphi'_{0}= - \frac{1}{2} u_{0}c_{1}.
\end{equation}
Replacing $\varphi'_{0}$ by Eq. ((\ref{b42}) we get
\begin{equation}\label{isot2}
\frac{k'_{a}p}{k_{d,0}} = \frac{1}{c_{1}}
e^{-\frac{u_{0}c_{1}}{2kT}}.
\end{equation}
For the first layer we denote
\begin{equation}\label{e}
\varepsilon=\frac{u_{0}}{kT},
\end{equation}
while for the second
\begin{equation}\label{e'}
\varepsilon'=\frac{u_{0}c_{1}}{kT}.
\end{equation}

\section{Adsorption isotherm and Critical behaviour}

The adsorption isotherm can be obtained from Eqs. (\ref{isot1})
and (\ref{isot2}). The result for $\theta$, that is the total
coverage ($\theta = c_1 + c_2$), is shown in Fig. 3. We only show
$\theta$ as a function of the chemical potential (the difference
with the representation $\theta $vs. $p$ ($\propto \alpha $)
amounts to a logarithmic scale change).

The staircase form is apparent. The first step corresponds to the
filling of the first layer, the second one to the filling of the
second layer, and so on. It is clear that if we consider more
layers the structure will persist, but with a progressive
collapse of the step size \cite{ry,rz}. It is of interest to
determine how the step size for different layers depends on the
system parameters, in particular on $\varepsilon$.  The analysis
indicates that, for the first layer, $\Delta \varphi $, the step's
size, as a function of $\varepsilon $, scales as $\Delta \varphi
\sim \varepsilon ^\nu$ with $\nu \approx 2.0$. A similar analysis
for the second step indicates that $\nu \approx 1.9$, showing a
decrease in step's size, in qualitative agreement with known data
\cite{ry,rz}.

It is also possible to analyze the behaviour of the order
parameters (that is $\Delta c_i = c_i^+ - c_i^-$) near the
critical point, considering their dependence on $\alpha_i$  (that
are related with the partial pressure, i.e. $\alpha_i \sim p$).
The characterization of such critical behaviour is certainly of
interest \cite{STAT}. In our case, from Eqs. (\ref{b2},\ref{b3})
and (\ref{b32},\ref{b42}) and near the critical point, we have
\begin{equation}
c_1^+-c_1^-=\frac{1}{(\varepsilon)^{1/2}}(\varepsilon-4)^{1/2}=
\frac{1}{2} \eta^{1/2}
\end{equation}
with $\eta = \varepsilon-4$ (with $\varepsilon \sim 4$); and
\begin{equation}
c_2^+-c_2^-=\frac{1}{(\varepsilon_0)^{1/2}}(\varepsilon-
\varepsilon_0)^{1/2}= \eta'^{1/2}
\end{equation}
where $\eta' = \varepsilon-\varepsilon_0$ (with $\varepsilon_0
\sim 1$). For both layers, we see that the critical behaviour, as
could be expected, corresponds to the typical mean field one with
a critical exponent 1/2.

\section{Density Profiles}

In this section we show how to get the form of the density
profile of a propagating interface with constant velocities
$v_{1}$ for the first layer and $v_{2}$ for the second layer.
They can be obtained considering the following change of variables
in Eqs. (\ref{sol1c}): $c_{1}(\zeta_{1})$; with $\zeta_{1}=\xi
-v_{1} \tau$ and $c_{2}(\zeta_{2})$; with $\zeta_{2}=\xi' -v_{2}
\tau$. Where $c_{1}(\zeta_{1})\rightarrow c_{1}$ when $\zeta_{1}
\rightarrow \infty$, $c_{1}(\zeta_{1})\rightarrow c_{3}$ when
$\zeta_{1} \rightarrow -\infty$, and also
$c_{2}(\zeta_{2})\rightarrow c_{5}$ when $\zeta_{2} \rightarrow
\infty$ $c_{2}(\zeta_{2})\rightarrow c_{7}$ when $\zeta_{2}
\rightarrow -\infty$

With the indicated change of variables, the partial differential
equations (\ref{sol1c}) transform into ordinary differential
equations
\begin{eqnarray}
-v_{1}\frac{dc_{1}}{d\zeta_{1}}&=& g_{1}(c_{1}) +
\frac{d}{d\zeta_{1}}(D(c_{1})\frac{dc_{1}}{d\zeta_{1}}), \\
 -v_{2}\frac{dc_{2}}{d\zeta_{2}}&=& g_{2}(c_{2}) +
\frac{d}{d\zeta_{2}}(D(c_{2})\frac{dc_{2}}{d\zeta_{2}}),
\end{eqnarray}
where
\begin{eqnarray}
g(c_{1})&=&\alpha_{1}(1-c_{1}) -c_{1} \exp(-\varepsilon c_{1}) \\
D(c_{1})&=& 1-\varepsilon c_{1} (1-c_{1}),
\end{eqnarray}
for the first layer, and
\begin{eqnarray}
g(c_{2})&=&\alpha_{2}(1-c_{2})c_{1} - c_{2}\exp(-\varepsilon
c_1 c_{2}), \\
D(c_{2})&=& 1-\varepsilon c_{2} (1-c_{2})c_{1},
\end{eqnarray}
for the second layer.

For convenience we assume that the concentration jump is
localized at $\zeta=0$. It is worth noting that with this change
of variables the coverage behaves as
$$c_{1}(\zeta_{1})=c_{1}^{-} - q x \zeta_{1}^{1/2},
\,\,\,\, \zeta_{1}\rightarrow +0,$$ and
$$c_{1}(\zeta_{1})=c_{1}^{+} - q x \zeta_{1}^{1/2},
\,\,\,\, \zeta_{1}\rightarrow -0,$$ with $q$ some
constant. Similarly, for the second layer
$$c_{2}(\zeta_{2})=c_{2}^{-} - a x \zeta_{2}^{1/2},\,\,\,\,
\zeta_{2}\rightarrow +0,$$ and
$$c_{2}(\zeta_{2})=c_{1}^{+} - a x \zeta_{2}^{1/2},
\,\,\,\,\ \zeta_{2}\rightarrow -0$$ with $a$ some constant.

The mass balance requires that the flux be the same on both sides
of the profiles moving with velocities $v_{1}$ and
$v_{2}$,respectively. Hence, we consider the equation
\begin{eqnarray}
D(c_{1}) \frac{dc_{1}}{d\zeta_{1}}_{\zeta_{1}\rightarrow +0} -
D(c_{1}) \frac{dc_{1}}{d\zeta_{1}}_{\zeta_{1}\rightarrow
-0}&=&-v_{1}(c_{1}^{+}-c_{1}^{-}), \\
D(c_{2}) \frac{dc_{2}}{d\zeta_{2}}_{\zeta_{2}\rightarrow +0} -
D(c_{2}) \frac{dc_{2}}{d\zeta_{2}}_{\zeta_{2}\rightarrow
-0}&=&-v_{2}(c_{2}^{+}-c_{2}^{-}).
\end{eqnarray}

We multiply by $D(c_{i})\frac{dc_{i}}{d\zeta_{i}},$ with $i=1,2$,
respectively. After integrating from $-\infty$ to $\infty$; we
obtain for the velocity of the first layer
\begin{eqnarray}
v_{1}= \frac{G(c_{3})-G(c_{1}^{+}) -
G(c_{1})+G_(c_{1}^{+})}{\int_{-\infty}^{\infty}
(\frac{dc_{1}}{d\zeta_{1}})^{2} D(c_{1}) d\zeta_{1}},
\end{eqnarray}
while for the second layer we have
\begin{eqnarray}
v_{2}= \frac{G(c_{7})-G(c_{2}^{+}) -
G(c_{5})+G_(c_{2}^{+})}{\int_{-\infty}^{\infty}
(\frac{dc_{2}}{d\zeta_{2}})^{2} D(c_{2}) d \zeta_{2}},
\end{eqnarray}
that in general is different from the velocity in the first layer.
Note that the propagation velocity goes to zero if the numerator
of the previous expressions goes to zero. This allows us to
represent the Maxwell conditions for the coexistence in
stationary conditions, for phases subject to first order
transitions
\begin{eqnarray}
G(c_{3}) - G(c_{1})&=& G(c_{1}^{+})- G(c_{1}^{-}) \\
G(c_{7}) - G(c_{5})&=& G(c_{2}^{+})- G(c_{2}^{-})
\end{eqnarray}

The profile in the first layer is described by the following
integral $$ \xi(c)=\frac{\int_{c}^{c_{1}^{-}} D(c)
dc}{[2x(G(c_{1})-G(c))]^{0.5}} $$ for $ c_{1}^{-}>c>c_{1}$, and
$$\xi(c)=-\frac{\int_{c_{1}^{+}}^{c}
D(c)dc}{[2x(G(c_{3})-G(c))]^{0.5}} $$ for $ c_{3}>c>c_{1}^{+}$.

For the second layer we have $$\xi'(c)=\frac{\int_{c}^{c_{2}^{-}}
D(c) dc}{[2x(G(c_{5})-G(c))]^{0.5}} $$ for $ c_{2}^{-}>c>c_{5}$,
and $$\xi'(c)=-\frac{\int_{c_{2}^{+}}^{c} D(c)
dc}{[2x(G(c_{7})-G(c))]^{0.5}} $$ for $ c_{7}>c>c_{2}^{+}$. For
$\zeta=0$ the coverage jumps at $c_{1}=c_{1}^{-}$ at
$c_{1}=c_{1}^{+}$, or in the second layer for
$c_{1}+c_{2}=c_{1}^{+}+c_{2}^{-}$ at $c_{1}+c_{2}=c_{1}^{+}+
c_{2}^{+}$.

Figure 5 shows an example of a typical form of the density
profiles. It is worth remarking here that the form of these
profiles is reminiscent of those found in the problem of layering
in wetting \cite{wett}. However, such an analogy requires further
study.

\section{Conclusions}

We have introduced a simple model able to describe qualitatively
multilayer adsorption and the formation of ordered structures. It
is an extension of an early model \cite{r1} describing a system
of a monolayer of adsorbed particles with lateral attractive
interactions leading to the formation of interfaces separating
phases with different local coverages. However, for sake of
simplicity, we have focused in the case of two layers. Such a
case show evidences indicating the occurrence of a ``cascade" of
first order phase transitions in the successive layers. Such a
result is in agreement with some experimental measurements
\cite{ry} as well as simulations \cite{rz}.

Lateral interactions play a key role in the whole picture. Also
the possibility of transfer interactions among different layers
can strongly affect the results, particularly the position of the
phase transition and the scaling of the step size $\Delta \alpha$
as a function of $\varepsilon $. The study of this exchange
effect will be done in a forthcoming work. Here we have found
that $\Delta \alpha \sim \varepsilon ^\nu$ with $\nu \approx 2.0$ for 
the first step while $\nu \approx 1.92$ for the second. We have also
found that the critical behaviour corresponds to a mean field
one, indicating that the analysis of critical properties requires
a more detailed study.

The study of  the fronts connecting different stationary states
\cite{RL}, for instance the dependence of the front velocity on
system's parameters, as well as the comparison of isotherm with
simulations and experiments \cite{sanluis}, is under way. Also,
the connection of this problem with the determination of more
realistic adsorption isotherm as well as with cases involving
reactions \cite{r3,r4} or layering problems in wetting
\cite{wett} will be the subject of further work. \\ \\

\bigskip\noindent{\small{\bf ACKNOWLEDGMENTS}}
The authors thank V. Grunfeld for a critical revision of the
manuscript. HSW thanks J.L.Riccardo for a fruitful discussion. We
also acknowledge partial provided by CONICET (grant PIP 4953/97),
Argentina.

\newpage

\section{Appendix}

In this appendix we show how to derive the macroscopic master
equation starting from a microscopic analysis. To reach such a
goal we closely follow the procedure of Ref.\cite{micros},
however with the adequate changes for the actual situation. The
basic step is to consider that the system is composed of a given
number of particles enclosed in a box and that this number is so
large that the local coverage, defined as the ratio between
occupied sites to the total number of sites, should not suffer an
appreciable change due to individual adsorption, desorption, or
diffusion processes.

\subsection{Microscopic approach to the system's evolution
equations}

For a box $j$ containing $n_{j}$ particles, the probability per
unit time that an adsorption or desorption occurs in the first
layer is proportional to the number of sites occupied or free in
the first and those free in the second layer. Hence, for the
first layer we have
\begin{equation}
\tilde{w}_{a}(n_{j})=w^{a} (N-n_{j})(1-n'_{j}/N'),
\end{equation}
\begin{equation}
\tilde{w}_{d}(n_{j})=w^{d}_{j} n_{j}(1-n'_{j}/N').
\end{equation}
Where $w^{a}=k_{a}p$ and $w^{d}_{j}=k_{d,0}\exp(U_{i}/k_{B}T)$,
and $N'$ is the maximum number of sites.

Here $\hat{w}_{j}^{\pm}$ the transition probability per unit time
for transitions between neighbor boxes (diffusion) is
proportional to $n_{j}$, the particle number in the box $j$, and
the fraction of free sites in the box towards which the transition
occurs
\begin{equation}
\tilde{w}_{j}^{\pm}=w_{j} ^{\pm}(1-n_{j \pm 1}/N)n_{j}.
\end{equation}
Where
\begin{equation}
w_{j}^{\pm}=\nu e^{\frac{U_{j}-U_{j\pm 1}}{k_{B}T}} \qquad
U_{j}<U_{j \pm 1},
\end{equation}
\begin{equation}
 w_{j}^{\pm}=\nu
\qquad U_{j}>U_{j \pm 1}.
\end{equation}

Similarly, for the second layer we can write
\begin{equation}
\tilde{w}'_{a}(n'_{j})=w'^{a} (N'-n'_{j})n_{j}/N,
\end{equation}
\begin{equation}
\tilde{w}'_{d}(n'_{j})=w'^{d}_{j} n'_{j}.
\end{equation}
Where $w'^{a}=k'_{a}p$, $w'^{d}_{j}=k'_{d,0} \exp(U'_{i}/k_{B}T)$,
and $N'$ is the maximum number of sites in the second layer ($N <
N'$).
\begin{equation}
\tilde{w}^{' \pm}_{j}={w}^{' \pm}_{j} (1-n'_{j \pm 1}/N')n'_{j}.
\end{equation}
Where
\begin{equation}
w_{j}^{' \pm}=\tilde{\nu}
 e^{\frac{U'_{j}-U'{j\pm 1}}{k_{B}T}} \qquad U'_{j}<U'_{j\pm 1},
\end{equation}
\begin{equation}
w_{j}^{' \pm}=\nu' \qquad U'_{j}>U'_{j\pm 1}.
\end{equation}

For the first layer we assumed $N=N_{max}>>1$, a number of
particles $n_{j}$ and $x_{j}$ sites with $j=1,\cdots ,m$. The
concentration is defined as $c_{j}=\frac{n_{j}}{N}$. For the
second layer we consider $N'=\sum n_{j} $, with $N'<N_{max}$ and
$N'>>1$. Here we have a number of particles $n'_{j}$ and $x_{j}$
sites with $j=1,\cdots, m$, and the concentration is defined as
$c'_{j}=\frac{n'_{j}}{N'}$.

We will try to obtain the master equation from the
multidimensional probability distribution
$p({n_{1},\cdots,n_{m}},t)$. Such a distribution corresponds to
the probability of finding $n_{1},\cdots,n_{m}$ particles in the
box located at  $x_{1},\cdots,x_{m}$, at time $t$. Let us see what
we obtain from such an analysis for the first and second layers.

\subsection{Equation of motion for the first layer}

The analysis of the time derivative of the probability
distribution $p({n_{1},\cdots,n_{m}},t)$ for the first layer gives
\begin{eqnarray}
\frac{\partial p(\{n_j\},t)}{\partial t }&=& w^{a} \sum_{j=1}^m
\frac{(N'-n'_j)}{N'}[(N-n_j +1) \hat{p}_j^{-}
-(N-n_j)p(\{n_j\},t)]+
 \nonumber \\ &+&\sum_{j=1}^{m} w^d_j
\frac{(N'-n'_{j})}{N'}[(n_j +1) \hat{p}_j^+ - n_j p(\{n_j\},t)]+
\nonumber \\ &+& \sum_{j=1}^m \sigma_j (n_j +1)
[(1-\frac{n_{j+1}-1}{N}) p^+_j + (1- \frac{n_{j-1}-1}{N})p^-_j]-
\nonumber \\ &-& \sum_{j=1}^{m} \sigma_j n_j
[2-\frac{n_{j+1}+n_{j-1}}{N}] p(\{n_j\},t) +\nonumber
\\ &+& \sum_{j=1}^{m} \gamma_j (n_j+1)
[(1-\frac{n_{j+1}-1}{N})p^{+}_{j}- (1-\frac{n_{j-1}-1}{N}) p^-_j]+
\nonumber \\
&+&\sum_{j=1}^{m}[\gamma_{j}n_{j}(\frac{n_{j+1}-n_{j-1}}{N})
p(\{n_j\},t)]. \label{MasterEq}
\end{eqnarray}
Remembering that
\begin{equation}
N'=\sum_{j=1}^{m} n_{j},
\end{equation}
the local coverage in the second layer is related to
\begin{equation}
c'_{j}=\frac{n'_{j}}{N'},
\end{equation}
where $N'>> 1$. These are the available sites for adsorption on
the second layer. Also $N >> 1$  and  the local coverage for the
first layer is $c_{j}=\frac{n_{j}}{N}$.

Assuming that $c_{j}$ changes only slightly with an adsorption or
desorption process, we can expand
\begin{equation}
\hat{p}^{\pm}\simeq p \pm \frac{1}{N}\frac{\partial{p}}{\partial
{c_{j}}}+\frac{1}{2N^{2}}\frac{\partial^{2}{p}}{\partial{
c_{j}^{2}}},
\end{equation}
hence,
\begin{equation}
p^{\pm}_{j} \simeq p \pm \frac{1}{N}[{\frac{\partial{p}}{\partial
{c_{j}}}-\frac{\partial{p}}{\partial{c_{j\pm 1}}}}]+
\frac{1}{N^{2}}[{\frac{1}{2} \frac{\partial^{2}{p}}{\partial
{c_{j}^{2}}}+\frac{1}{2} \frac{\partial^{2}{p}}{\partial^{2}
{c_{j\pm 1}}}- \frac{\partial^{2}{p}}{\partial{c_{j}}\partial
{c_{j\pm 1}}}}].
\end{equation}
Replacing this into Eq. (\ref{MasterEq}) we get
\begin{eqnarray}
\frac{\partial p(\{c_{j}\},t)}{\partial t}&=& -\sum_{j=1}^{m} w^a
(1-c'_j) \frac{\partial }{\partial c_j} [(1-c_j) p] + \frac{1}{2N}
\sum_{j=1}^{m} w^a (1-c'_j) \frac{\partial^2 } {\partial c_j^2}
[(1-c_j) p]-
 \nonumber \\ &-& \sum_{j=1}^{m} w_{j}^{d} (1-c'_j)
\frac{\partial }{\partial c_j} [-c_j p] + \frac{1}{2N}
\sum_{j=1}^{m} w_{j}^d (1-c'_j) \frac{\partial ^2}{\partial c_j^2}
[c_j p]- \nonumber \\ &-& \sum_{j=1}^{m} \frac{\partial }{\partial
c_j} [(\sigma_{j+1}c_{j+1}+ \sigma_{j-1} c_{j-1}-2c_j \sigma_j)
(1-c_j) p]- \nonumber \\ &-& \sum_{j=1}^{m} \frac{\partial
}{\partial c_j} [(c_{j+1}+c_{j-1}-2c_j) \sigma_j c_j p]+ \nonumber
\\ &+& \frac{1}{2N} \sum_{j=1}^{m} \frac{\partial ^2}{\partial
c_j^2} \{ [(\sigma_{j+1} c_{j+1} + \sigma_{j-1} c_{j-1}-2 c_j
\sigma_j )(1-c_j)-  \nonumber \\ & &  \,\,\,\,\,\,\,\,\,\ -
(c_{j+1}+c_{j-1} - 2 c_j) \sigma_j c_j ] p \}-\nonumber \\
&-&\frac{1}{N}\sum_{j=1}^{m} \frac{\partial }{\partial c_j} [
\frac{\partial }{\partial c_{j+1}}(1-c_{j+1})+ \frac{\partial }
{\partial c_{j-1}} (1-c_{j-1})-  \nonumber \\ & &
\,\,\,\,\,\,\,\,\,\ -2 \frac{\partial } {\partial c_j} (1-c_j)]
(\sigma_j c_j p)- \nonumber \\ &-& \sum_{j=1}^{m} \frac{\partial
}{\partial c_j} [[ \gamma_j c_j (c_{j+1}-c_{j-1})+ (c_j-1)(
\gamma_{j+1}c_{j+1}- \gamma_{j-1} c_{j-1}) ] p]- \nonumber \\ &-&
\frac{1}{2N}\sum_{j=1}^{m} \frac{\partial ^2}{\partial
c_j^2}[[\gamma_j c_j (c_{j+1}-c_{j-1})+ (1-c_j) (\gamma_{j+1}
c_{j+1}- \gamma_{j-1}c_{j-1})]p]- \nonumber \\ &-& \frac{1}{N}
\sum_{j=1}^{m} \frac{\partial }{\partial c_j} [\frac{\partial
}{\partial c_{j+1}} (1-c_{j+1}) - \frac{\partial }{\partial
c_{j-1}} (1-c_{j-1})] (\gamma_j c_j p).
\end{eqnarray}

As indicated above, the size of an individual box is much smaller
than the characteristic length scale of the spatial patterns.
Therefore, both coverage, $c_{j}$ and $c'_{j}$, do not
significantly change between neighbor boxes, and we can assume
that the coverage $c(x)$ is continuous. Hence we consider a
continuous version of the equation in terms of $c(x)$. These
transformation also transform the multidimensional distribution
$p(\{c_{j}\},t)$ into the functional $p(c(x),t)$. The evolution
equations for such a functional become
\begin{eqnarray}
\frac{\partial{p}}{\partial{t}}&=& - \int dx \frac{\delta}{\delta
c(x)}{[(1-c') [w^{a}(1-c)-w^{d}c]+ 2 l_{o} \frac{\partial{(\gamma
c(1-c))}} {\partial{x}}]p(c(x),t)}  \nonumber \\
&-&l_{o}^{2}\int dx \frac{\delta}{\delta c(x)}[(1-c)
\frac{\partial^{2}{(\sigma c)}}{\partial{x}^{2}}+ \sigma c
\frac{\partial^{2}{c}}{\partial {x}^{2}}]p(c(x),t) \nonumber \\
&+&\frac{1}{2\mu} \int dx \frac{\delta^{2}}{\delta c(x)^{2}}
{[(1-c')(w^{a}(1-c)+w^{d} c)+ l_{o}^{2}(1-c)
\frac{\partial^{2}{(\sigma c)}}{\partial{x}^{2}}- \sigma l_{o}^{2}
c \frac{\partial^{2}{c}}{\partial {x}^{2}}]p(c(x),t)} \nonumber
\\
&+& \frac{l_{o}}{2\mu} \int dx \frac{\delta^{2}}{\delta
c(x)^{2}}{[[2 \gamma c \frac{\partial {c}}{\partial{x}}-2(1-c)
\frac{\partial{(\gamma c)}}{\partial{x}}] p(c(x),t)]} \nonumber
\\
&-& \frac{l_{o}^{2}}{\mu} \int dx \frac{\delta}{\delta c(x)}
\frac{\partial^{2}{}}{\partial{x}^{2}}[\frac{\delta}{\delta c(x)}
(1-c(x))]{\sigma c p(c(x),t)} \nonumber \\ &-& \frac{l_{o}}{\mu}
\int dx \frac{\partial{}}{\partial{x}} ((\frac{\delta}{\delta
c(x)})^{2}\gamma c(1-c) p(c(x),t)).
\end{eqnarray}
Here we have introduced the parameter $\mu=\frac{N}{l_{o}}$,
where  $l_{o}$ is the typical size of each box. This value of
$\mu$ give us the number of lattice sites per unit area. .

The coefficients $\sigma$ and $\gamma$ in the previous equation
represent functions of the coordinate $x$ given by
\begin{equation}
\sigma_{j}=\frac{w^{+}_{j}+w^{-}_{j}}{2},
\end{equation}
and
\begin{equation}
\gamma_{j}=\frac{w^{+}_{j}-w^{-}_{j}}{2}.
\end{equation}
Here
\begin{equation}
w^{+}_{j}=\nu e^{\frac{U_{j}-U_{j\pm1}}{k_{B}T}} \qquad
U_{j}<U_{j\pm1},
\end{equation}
and
\begin{equation}
w^{-}_{j}=\nu \qquad U_{j}>U_{j\pm1}.
\end{equation}
Hence,
\begin{equation}
\sigma(x)=\frac{\nu}{2}[1+ e^{\frac{-l_{o}}{K_{B}T}|
\frac{\partial{U}}{\partial{x}}|}],
\end{equation}
and
\begin{equation}
\gamma(x)=-\frac{\nu}{2}[1- e^{\frac{-l_{o}}
{K_{B}T}|\frac{\partial{U}}{\partial{x}}|}]\,\aleph ,
\end{equation}
with $\aleph $ indicating the sign of $ \frac{\partial
U}{\partial x}$.

Considering the limit $l_{o} \to 0$, we get
\begin{equation}
lim_{l_{o} \to 0} (\gamma(x) l_{o}^{2}) = D,
\end{equation}
where $D$ is the diffusion constant
\begin{equation}
D=lim_{l_{o} \to 0} (\nu l_{o}^{2}),
\end{equation}
and
\begin{equation}
lim_{l_{o} \to 0} [\gamma(x) l_{o}]= lim_{l_{o} \to 0}
[-\frac{\nu l_{o}^{2}}{2K_{B}T} \frac{\partial{U}}{\partial{x}}]=
-\frac{1}{2}\frac{D}{K_{B}T}\frac{\partial{U}}{\partial{x}}.
\end{equation}

Introducing the functional differential operator
\begin{equation}
\widehat{A}(x)=\frac{\delta}{\delta c(x)},
\end{equation}
considering the evolution equation in the limit $l_{o}\to 0$, and
using the previous equations we obtain
\begin{eqnarray}
\label{a36} \frac{\partial{p}}{\partial{t}}&=& - \int dx
\hat{A}(x)[{(1-c') [w^{a}(1-c)-w^{d}c]+ \frac{\partial{}}{\partial
{x}}(\frac{D}{K_{B}T} \frac{\partial{U}}{\partial{x}} c(1-c))+ D
\frac{\partial^{2}{c}}{\partial{x}^{2}}}]p+\nonumber \\
&+&\frac{1}{2\mu}\int dx \widehat{A}^{2}(x)[(1-c')[w^{a}(1-c)+
w^{d}c]+D(1-2c)\frac{\partial^{2}{c}}{\partial{x}^{2}}-\nonumber
\\
&-& \frac{D c}{K_{B}T}\frac{\partial{U}}{\partial{x}}
\frac{\partial{c}}{\partial
{x}}+(1-c)\frac{\partial{}}{\partial{c}}
(\frac{D c}{K_{B}T}\frac{\partial{U}}{\partial{x}})]p-\nonumber \\
&-&\frac{D}{\mu} \int dx
[\widehat{A}(x)\frac{\partial^{2}{}}{\partial{x}^{2}}
(\widehat{A}(x)(1-c))c-\frac{1}{2}\frac{\partial{}}{\partial
{x}}(\widehat{A}(x)^{2}) \frac{1}{K_{B}T}c(1-c)\frac{\partial
{U}}{\partial{x}}]p.
\end{eqnarray}

We now consider those terms inversely proportional to $\mu$ and
independent of the potential $U$,
\begin{eqnarray}
F_{d}&=&\frac{D}{\mu}\int dx [\hat{A}^{2}(x)[\frac{1}{2}(1-2c)
\frac{\partial^{2}{c}}{\partial{x}^{2}}]-\hat{A}(x)
\frac{\partial^{2}{}}{\partial {x}^{2}}[\hat{A}(x)(1-c)]c]p.
\end{eqnarray}
This expression can be further transformed into
\begin{eqnarray}
F_{d}&=&\frac{D}{\mu}\int dx [\hat{A}^{2}(x)[\frac{1}{2}(1-2c)
\frac{\partial^{2}{c}}{\partial{x}^{2}}+
c\frac{\partial^{2}{c}}{\partial{x}^{2}}]+2\hat{A}(x)\frac{\partial
{\widehat{A}(x)}}{\partial{x}} \frac{\partial{c}}{\partial{x}}c-
\widehat{A}(x)\frac{\partial^{2}{}}{\partial{x}^{2}}
[\widehat{A}(x)(1-c)c]]p \nonumber \\
&=&\frac{D}{\mu}\int dx
[\widehat{A}^{2}(x)[\frac{1}{2}\frac{\partial^{2}{c}}{\partial{x}^{2}}]+
 c\frac{\partial{\widehat{A}^{2}(x)}}{\partial{x}}
\frac{\partial{c}}{\partial{x}}-\frac{1}{2}\frac{\partial^{2}{}}{\partial
{x}^{2}}[\widehat{A}^{2}(x)(1-c)c+ [\frac{\partial
{\widehat{A}}}{\partial{x}}]^{2}(1-c)c]]p\nonumber \\
&=&\frac{D}{\mu}\int dx
[\widehat{A}^{2}(x)[\frac{1}{2}\frac{\partial^{2}{c}}{\partial{x}^{2}}]+
\frac{1}{2} \frac{\partial{\widehat{A}(x)}}{\partial{x}}
\frac{\partial{c}}{\partial{x}}+[\frac{\partial{\widehat{A}}}
{\partial{x}}](1-c)c]p,
\end{eqnarray}
\begin{equation}
F_{d}=\frac{D}{\mu}\int dx (\frac{\partial{\widehat{A}}}{\partial
{x}})^{2}c (1-c) p,
\end{equation}
\begin{equation}
F_{d}=\frac{D}{\mu}\int \int dx dy
(\frac{\partial{\widehat{A}(x)}}{\partial{x}}) (\frac{\partial
{\widehat{A}(y)}}{\partial{y}})c(x) (1-c(x))\delta(x-y) p,
\end{equation}
\begin{equation}
F_{d}=\frac{1}{2\mu}\int \int dx dy \widehat{A}(x)\widehat{A}(y)
\frac{\partial^{2}{}}{\partial{x} \partial{y}}[2 D
c(x)(1-c(x))\delta(x-y)] p.
\end{equation}

Now we consider in Eq. (\ref{a36}) the terms of order
$O(\mu^{-1})$, that are proportional to the gradient of the
potential $U$. We find that such terms cancel each other, i.e.
\begin{eqnarray}
F&=&\frac{D}{2\mu k_{B}T}\int
dx[{\widehat{A}^{2}(x)[-\frac{\partial{U}}{\partial{x}}c
\frac{\partial {c}}{\partial{x}}+(1-c)\frac{\partial{}}{\partial
{x}}(c\frac{\partial{U}}{\partial{x}})]+
\frac{\partial{}}{\partial{x}}(\widehat{A}^{2}(x)
c(1-c)\frac{\partial{U}}{\partial{x}})}]p\nonumber \\
&=&\frac{D}{2\mu k_{B}T}\int dx
\widehat{A}^{2}(x)[-\frac{\partial U}{\partial x}c \frac{\partial
c}{\partial x}+(1-c)\frac{\partial }{\partial x}(c \frac{\partial
U}{\partial x})- \frac{\partial }{\partial x}(c(1-c)
\frac{\partial U}{\partial x})]p=0.
\end{eqnarray}

Using the previous results the evolution equation can be written
as
\begin{eqnarray}
\frac{\partial{p}}{\partial{t}}&=&-\int dx \frac{\delta}{\delta
c(x)}[ {w^{a}(1-c')(1-c)-w^{d}(1-c')c+
\frac{D}{k_{B}T}\frac{\partial{}}{\partial
{x}}(c(1-c))\frac{\partial{U}}{\partial{x}})+
D\frac{\partial^{2}{c}}{\partial{x}^{2}}]p}+\nonumber \\
&+&\frac{1}{2\mu} \int \int dx dy \frac{\delta^{2}}{\delta c (x)
\delta c(y)}[[(1-c') [w^{a}(1-c)+w^{d} c]] \delta(x-y)+\nonumber \\
&+&\frac{\partial^{2}{}}{\partial{x}
\partial{y}} (2Dc (1-c) \delta(x-y))]p.
\end{eqnarray}

The last expression corresponds to the functional Fokker-Planck
equation for the probability distribution functional $p(c(x),t)$.
From the theory of stochastic processes, this Fokker-Planck
equation is equivalent to a stochastic partial differential
\cite{barza} (Langevin-like) equation for the fluctuating field
$c(x,t)$. It has the form
\begin{eqnarray}
\frac{\partial{c}}{\partial{t}}&=&w^{a}(1-c)(1-c')-w^{d}
c(1-c')+\frac{D}{k_{B}T} \frac{\partial{}}{\partial{x}}[c (1-c)
\frac{\partial{U}}{\partial{x}}] +\nonumber \\
&+&D\frac{\partial{} ^{2} c}{\partial{x}^{2}}+
\frac{1}{\mu^{\frac{1}{2}}}[w^{a}(1-c')(1-c)]^{\frac{1}{2}}
f_{a}(x,t)+
\frac{1}{\mu^{\frac{1}{2}}}[w^{d}c(1-c')]^{\frac{1}{2}}
f_{d}(x,t)+\nonumber \\
&+& \frac{1}{\mu^{\frac{1}{2}}} \frac{\partial{}}{\partial {x}}
[(2Dc(1-c))^{\frac{1}{2}} f(x,t)].
\end{eqnarray}
Here $f_{a}(x,t)$,$f_{d}(x,t)$ and $f(x,t)$, are independent
noise sources, with intensity one according to the Ito
interpretation. Those three noise sources correspond to internal
noises associated to adsorption, desorption and diffusion
respectively. When studying the equation in the macroscopic limit,
such noise sources are neglected.

\subsection{Equation of motion for the second layer}

Next, we analyze the time derivative of the probability density
$p(\{n'_{j}\},t)$ for the second layer. With similar arguments as
before we get
\begin{eqnarray}
\frac{\partial p(\{n'_j\},t)}{\partial t} &=& w^{'a}
\sum_{j=1}^{m} \frac{n_j}{N}(N'-n'_j+1) \tilde{p}_j^{'-}-(N'-n'_j)
p'(\{n'_{j}\},t)+ \nonumber \\ &+& \sum_{j=1}^{m}
w'^{d}_{j}[(n'_j+1) \tilde{p}_j^{'+} -n'_j p'(\{n_{j}\},t)]+
\nonumber \\ &+& \sum_{j=1}^{m} [\sigma'_j(n'_j+1) [(1-
\frac{n'_{j+1}-1}{N'}) p_j^{'+} + (1-\frac{n'_{j-1}-1}{N'})
p_j^{'-}]]- \nonumber \\ &-& \sum_{j=1}^{m} [\sigma'_{j} n'_{j}(2-
\frac{n'_{j+1}+n'_{j-1}} {N'}) p'(\{n'_{j}\},t)]+ \nonumber \\ &+&
\sum_{j=1}^{m} \gamma'_{j}(n'_{j}+1) [(1- \frac{n'_{j+1}-1}{N'})
p_j^{'+} - (1-\frac{n'_{j-1}-1}{N'}) p_j^{'-}]+ \nonumber
\\ &+& \sum_{j=1}^{m} [\gamma'_{j} n_j
(\frac{n'_{j+1}-n'_{j-1}}{N'}) p'(\{n'_{j}\},t)]. \label{seco}
\end{eqnarray}
We should remember that
\begin{eqnarray}
N'=\sum_{j=1}^{m} n_{j},
\end{eqnarray}
and the coverage of the second layer is related to
\begin{eqnarray}
c'_{j}=\frac{n'_{j}}{N'},
\end{eqnarray}
where $N'>> 1$, $N'<  N_{max}$; and $N'$ are the available sites
on the second layer (that is the occupied sites in the first
layer).

As in the case of the first layer we approximate
\begin{eqnarray}
\tilde{p'}^{\pm}\approx p'\pm
\frac{1}{N'}\frac{\partial{p'}}{\partial {c'_{j}}}+
\frac{1}{2N^{2}}\frac{\partial^{2}{p'}}{\partial {c'_{j}}^{2}},
\end{eqnarray}
and
\begin{eqnarray}
p'^{\pm}_{j} \approx p'+
\frac{1}{N'}[{\frac{\partial{p'}}{\partial
{c'_{j}}}-\frac{\partial{p'}}{\partial{c'_{j\pm1}}}}]+
\frac{1}{N^{2}}[\frac{1}{2} \frac{\partial^{2}{p'}}{\partial
{c'_{j}}^{2}}+\frac{1}{2} \frac{\partial^{2}{p'}}{\partial
{c'_{j\pm
1}}^{2}}+\frac{\partial{p'}^{2}}{\partial{c'_{j}}\partial{c'_{j\pm
1}}}].
\end{eqnarray}

Replacing into Eq. (\ref{seco}) we obtain
\begin{eqnarray}
\frac{\partial{p(\{c'_{j}\},t)}}{\partial{t}}&=&
-\sum_{j=1}^{m}c_{j}w^{' a} \frac{\partial{}}{\partial
{c'_{j}}}[(1-c'_{j})p']+ \frac{1}{2N'}\sum_{j=1}^{m} w^{'
a}c_{j}\frac{\partial^{2}{}} {\partial
{c^{' 2}_{j}}}[(1-c'_{j})p']- \nonumber \\
&-&\sum_{j=1}^{m}\frac{\partial{}}{\partial{c'_{j}}}(w^{' d}_{j}
c'_{j}p')+
\frac{1}{2N'}\sum_{j=1}^{m}\frac{\partial^{2}{}}{\partial {c^{'
2}_{j}}} (w^{' d}_{j} c'_{j}p')- \nonumber \\
&-&\sum_{j=1}^{m}\frac{\partial{}}{\partial{c'_{j}}}
[\sigma'_{j} c'_{j}p'(c'_{j+1}-2c'_{j}+c'_{j-1})]+\nonumber \\
&+&\frac{1}{2N'}\sum_{j=1}^{m}\frac{\partial^{2}{}}{\partial
{c'_{j}}^{2}}
[(\sigma'_{j+1}c'_{j+1}+\sigma'_{j-1}c'_{j-1}-2c'_{j}\sigma'_{j})
(1-c'_{j})- \nonumber \\ & & \,\,\,\,\,\,\,\,\,\ -
(c'_{j+1}+c'_{j-1}-2c'_{j})\sigma'_{j}c'_{j}]p'-\nonumber \\
&-&\frac{1}{N'}\sum_{j=1}^{m}\frac{\partial{}}{\partial{c'_{j}}}
[\frac{\partial{}}{\partial{c'_{j+1}}}(1-c'_{j+1})+
\frac{\partial{}}{\partial{c'_{j-1}}}
(1-c'_{j-1})-2\frac{\partial{}}{\partial
{c'_{j}}}(1-c'_{j})][\sigma'_{j}c'_{j}p']-\nonumber \\
&-&\sum_{j=1}^{m} \frac{\partial{}}{\partial
{c'_{j}}}[{\gamma'_{j}c'_{j}(c'_{j+1}-c'_{j-1})+
(c'_{j}-1)(\gamma'_{j+1}c'_{j+1}-\gamma'_{j-1}c'_{j-1})}]p'-\nonumber
\\
&-& \frac{1}{2N'}\sum_{j=1}^{m} \frac{\partial{}^{2}}{\partial
{c'_{j}}^{2}} [{\gamma'_{j}c'_{j}(c'_{j+1}-c'_{j-1})+
(1-c'_{j})(\gamma'_{j+1}c'_{j+1}-\gamma'_{j-1}c'_{j-1})}]p'-
\nonumber \\
&-& \frac{1}{N'}\sum_{j=1}^{m} \frac{\partial{}}{\partial{c'_{j}}}
[{\frac{\partial{}}{\partial{c'_{j+1}}}(1-c'_{j+1})-
\frac{\partial{}}{\partial{c'_{j-1}}}
(1-c_{j-1})}](\gamma'_{j}c'_{j}p').
\end{eqnarray}

As for the first layer we consider a uniform coverage implying
\begin{eqnarray}
c(x_{j}) \to c(x),
\end{eqnarray}
\begin{eqnarray}
c'(x_{j}) \to c'(x).
\end{eqnarray}
Also
\begin{eqnarray}
p(\{c_{j}\},t) \to p(c(x),t),
\end{eqnarray}
\begin{eqnarray}
p'(\{c'_{j}\},t) \to p(c'(x),t).
\end{eqnarray}

The continuous version of the equation of motion, assuming the
uniform coverage of the system as before, is again a functional
equation for $p(c'(x),t)$. This equation is
\begin{eqnarray}
\frac{\partial p(c'(x),t)}{\partial t} &=& - \int dx \frac{\delta
}{\delta c'(x)}[cw^{'a}(1-c')-w^{'d}c'+ 2 l_{o} \frac{\partial
(\gamma' c'(1-c'))}{\partial x}]p(\{c'(x)\},t)- \nonumber \\
&-&l_{o}^{2}\int dx \frac{\delta }{\delta c'(x)}[[(1-c')
\frac{\partial ^2 (\sigma' c')}{\partial x^2}+ \sigma' c'
\frac{\partial ^2 c'}{\partial x^2}] p(\{c'(x)\},t)]+ \nonumber
\\
&+& \frac{1}{2\mu'} \int dx \frac{\delta ^2}{\delta c'(x)^2}
[[(1-c') w^{'a} c + w^{'d} c' + l_{o}^2 (1-c') \frac{\partial ^2
(\sigma' c')}{\partial x^2}- \nonumber \\ && \,\,\,\,\,\, -
\sigma' l_{o}^2 c \frac{\partial ^2 c'}{\partial x^2}]
p(\{c'(x)\},t)]+ \nonumber \\ &+& \frac{l_{o}}{2\mu'} \int dx
\frac{\delta ^2}{\delta c'(x)^2}[[2 \gamma' c' \frac{\partial
c'}{\partial x}-2(1-c') \frac{\partial (\gamma' c')}{\partial x}]
p(\{c'(x)\},t)]- \nonumber \\ &-& \frac{l_{o}^{2}}{\mu'} \int dx
\frac{\delta }{\delta c'(x)} \frac{\partial ^2}{\partial x^2}
(\frac{\delta }{\delta c'(x)} (1-c'))[\sigma' c' p(\{c'(x)\},t)]-
\nonumber \\ &-& \frac{l_{o}}{\mu'} \int dx \frac{\partial
}{\partial x} ([\frac{\delta }{\delta c'(x)}]^2)(\gamma' c'(1-c')
p([c'(x)],t)).
\end{eqnarray}

Again as in the first layer case, we have introduced here the
parameter $\mu'=\frac{N'}{l_{o}}$, where $l_{o}$ is the size of
each box.  The value of $\mu'$ gives the number of sites per unit
of area.

The coefficients $\sigma'$ y $\gamma'$ in the last equation
differ from those in the first layer by a factor of  $\mu'$.
Hence,
\begin{equation}
\sigma'(x)=\frac{\nu'}{2}[1+
e^{\frac{-l_{o}}{K_{B}T}|\frac{\partial{U}}{\partial{x}}|}],
\end{equation}
\begin{equation}
\gamma'(x)=-\frac{\nu'}{2}[1- e^{\frac{-l_{o}}
{K_{B}T}|\frac{\partial U}{\partial x}|}]\, \,\aleph ,
\end{equation}
where $\aleph $, as before, indicate the sign of $ \frac{\partial
U}{\partial x}$.

In the limit $l_{o} \to 0$, we get
\begin{equation}
lim_{l_{o} \to 0} (\gamma'(x) l_{o}^{2}) =D',
\end{equation}
where $D'$ is the diffusion constant on the second layer,
\begin{equation}
D'=lim_{l_{o} \to 0} (\nu' l_{o}^{2}),
\end{equation}
and
\begin{equation}
lim_{l_{o} \to 0} [\gamma'(x) l_{o})]= lim_{l_{o} \to 0}
[-\frac{\nu' l_{o}^{2}}{2K_{B}T} \frac{\partial{U}}{\partial{x}}]=
-\frac{1}{2}\frac{D'}{K_{B}T}\frac{\partial{U}}{\partial{x}}.
\end{equation}
We will adopt $D'\simeq D$, as they differ by a constant.

We introduce now the differential operator
\begin{equation}
\widehat{B}(x)=\frac{\delta}{\delta c'(x)}.
\end{equation}
Introducing it into the evolution equation and considering the
limit $l_{o}\to 0$, we get
\begin{eqnarray}\label{a63}
\frac{\partial{p}}{\partial{t}}&=& - \int dx \hat{B}(x)[[c
w'^{a}(1-c')-w'^{d}c']+ \frac{\partial{}}{\partial
{x}}(\frac{D}{K_{B}T} \frac{\partial{U}}{\partial{x}} c'(1-c'))+
\nonumber \\ &+& D
\frac{\partial^{2}{c'}}{\partial{x}^{2}}]p(c'(x),t)+ \nonumber \\
&+&\frac{1}{2\mu'}\int dx \widehat{B}^{2}(x)[[cw'^{a}(1-c')+
w'^{d}c']+D(1-2c')\frac{\partial^{2}{c'}}{\partial{x}^{2}}-
\nonumber \\ &-& \frac{D
c'}{K_{B}T}\frac{\partial{U}}{\partial{x}}
\frac{\partial{c'}}{\partial
{x}}+(1-c')\frac{\partial{}}{\partial{c'}} (\frac{D
c'}{K_{B}T}\frac{\partial{U}}{\partial{x}})]p(c'(x),t)- \nonumber
\\ &-&\frac{D}{\mu'} \int dx
[\widehat{B}(x)\frac{\partial^{2}{}}{\partial{x}^{2}}
(\widehat{B}(x)(1-c'))c' -\nonumber \\ & &
\,\,\,\,\,\,-\frac{1}{2}\frac{\partial{}}{\partial
{x}}(\widehat{B}(x)^{2}) \frac{1}{K_{B}T}c'(1-c')\frac{\partial
{U}}{\partial{x}}]p(c'(x),t).
\end{eqnarray}

As for the previous first layer case, we consider the terms
inversely proportional to $\mu'$ and those independent of the
potential $U$.
\begin{eqnarray}
F'_{d}&=&\frac{D}{\mu'}\int dx [\hat{B}^{2}(x)[\frac{1}{2}(1-2c')
\frac{\partial^{2}{c'}}{\partial{x}^{2}}] - \hat{B}(x)
\frac{\partial^{2}{}}{\partial {x}^{2}}[\hat{B}(x)(1-c')c']]
p(c'(x),t).
\end{eqnarray}
This expression can be transformed into
\begin{eqnarray}
F'_{d}&=&\frac{D}{\mu'}\int dx [\hat{B}^{2}(x)[\frac{1}{2}(1-2c')
\frac{\partial^{2}{c'}}{\partial{x}^{2}}+
c'\frac{\partial^{2}{c'}}{\partial{x}^{2}}]+2\hat{B}(x)\frac{\partial
{\widehat{B}(x)}}{\partial{x}} \frac{\partial{c'}}{\partial{x}}c'-
\nonumber \\ && \,\,\,\,\,\,\,\,\,\ -
\widehat{B}(x)\frac{\partial^{2}{}}{\partial{x}^{2}}
[\widehat{B}(x)(1-c')c']]p \nonumber \\ &=&\frac{D}{\mu'}\int dx
[\widehat{B}^{2}(x)[\frac{1}{2}\frac{\partial^{2}{c'}}{\partial{x}^{2}}]+
 c'\frac{\partial{\widehat{B}^{2}(x)}}{\partial{x}}
\frac{\partial{c'}}{\partial{x}}-\frac{1}{2}\frac{\partial^{2}{}}{\partial
{x}^{2}}[\widehat{B}^{2}(x)(1-c')c'+ \nonumber \\ & &
\,\,\,\,\,\,\,\,\,\  + [\frac{\partial
{\widehat{B}}}{\partial{x}}]^{2}(1-c')c']p]\nonumber \\
&=&\frac{D}{\mu'}\int dx
[\widehat{B}^{2}(x)[\frac{1}{2}\frac{\partial^{2}{c'}}{\partial{x}^{2}}]+
\frac{1}{2} \frac{\partial{\widehat{B}(x)}}{\partial{x}}
 \frac{\partial{c'}}{\partial{x}}+
 [\frac{\partial{\widehat{B}}}{\partial{x}}](1-c')c']p,
\end{eqnarray}
\begin{equation}
F'_{d}=\frac{D}{\mu'}\int dx
(\frac{\partial{\widehat{B}}}{\partial {x}})^{2}c'(1-c') p,
\end{equation}
\begin{equation}
F'_{d}=\frac{D}{\mu'}\int \int dx dy
(\frac{\partial{\widehat{B}(x)}}{\partial{x}}) (\frac{\partial
{\widehat{B}(y)}}{\partial{y}})c'(x) (1-c'(x))\delta(x-y) p,
\end{equation}
\begin{equation}
F'_{d}=\frac{1}{2\mu'}\int \int dx dy \widehat{B}(x)\widehat{B}(y)
\frac{\partial^{2}{}}{\partial{x} \partial{y}}[2 D
c'(x)(1-c'(x))\delta(x-y)] p.
\end{equation}

Now we consider in Eq. (\ref{a63}) the terms of order
$O(\mu^{-1})$, that are proportional to the gradient of the
potential $U$. We find that such terms cancel each other, i.e.
\begin{eqnarray}
F'&=&\frac{D}{2\mu' k_B T}\int dx
[\widehat{B}^2(x)[-\frac{\partial U}{\partial x} c' \frac{\partial
c'}{\partial x} + (1-c') \frac{\partial }{\partial
x} (c'\frac{\partial U}{\partial x})] \nonumber \\
&& \,\,\,\,\,\, + \frac{\partial }{\partial x}(\widehat{B}^2(x)
c'(1-c') \frac{\partial U}{\partial x})]p \nonumber \\
&=& \frac{D}{2\mu' k_B T} \int dx \widehat{B}^2(x)
[-\frac{\partial U}{\partial x} c' \frac{\partial c'}{\partial x}
+ (1-c') \frac{\partial }{\partial x} (c' \frac{\partial
U}{\partial x}) \nonumber \\
& & \,\,\,\,\,\, - \frac{\partial }{\partial x} (c'(1-c')
\frac{\partial U}{\partial x})]p=0.
\end{eqnarray}

Taking into account our previous results, the evolution equation
becomes
\begin{eqnarray}
\frac{\partial{p}}{\partial{t}}&=&-\int dx \frac{\delta}{\delta
c'(x)} [w'^{a}(1-c')c-w'^{d}c' \nonumber \\
&&  \,\,\,\,\,\,\, + \frac{D}{k_{B}T}\frac{\partial{}}{\partial
{x}}(c'(1-c'))\frac{\partial{U}}{\partial{x}})+
D\frac{\partial^{2}{c'}}{\partial{x}^{2}}]p+\nonumber \\
&+&\frac{1}{2\mu'} \int \int dx dy \frac{\delta^{2}}{\delta c'(x)
\delta c'(y)}[[(1-c')w'^{a}c+w'^{d} c'] \delta(x-y)
\nonumber \\
&& \,\,\,\,\,\ + \frac{\partial^{2}{}}{\partial{x}
\partial{y}} (2Dc'(1-c')\delta(x-y))]p.
\end{eqnarray}
The last equation is the functional Fokker-Planck equation for
the  probability distribution  $p(c'(x),t)$ for the second layer.
Similarly to the discussion for the first layer, we can obtain
the related stochastic partial differential equation
\begin{eqnarray}
\frac{\partial{c'}}{\partial{t}}&=&w'^{a}c(1-c')-w'^{d}
c'+\frac{D}{k_{B}T} \frac{\partial{}}{\partial{x}}[c'(1-c')
\frac{\partial{U}}{\partial{x}}] +\nonumber \\
&+&D\frac{\partial{} ^{2} c'}{\partial{x}^{2}}+
\frac{1}{\mu^{\frac{1}{2}}}[w'^{a}(1-c')c]^{\frac{1}{2}}
f'_{a}(x,t)+ \frac{1}{\mu'^{\frac{1}{2}}}[w'^{d}c']^{\frac{1}{2}}
f'_{d}(x,t)+\nonumber \\
&+& \frac{1}{\mu'^{\frac{1}{2}}} \frac{\partial{}}{\partial {x}}
[(2Dc'(1-c'))^{\frac{1}{2}} f'(x,t)].
\end{eqnarray}
Here, as before, $f'_{a}(x,t)$,$f'_{d}(x,t)$ and $f'(x,t)$, are
independent noise sources, with intensity one according to the Ito
interpretation. These three noise sources correspond to internal
noises associated to adsorption, desorption and diffusion
respectively, but now in the second layer. Also, when studying
the equation in the macroscopic limit, such noise sources are
neglected.

The previous analysis leads us to Eqs.(\ref{sol1c}) and
(\ref{sol2c}).

\newpage

\begin{figure}[tbp]
\caption{Sketch of the different processes included in the model.}
\end{figure}

\begin{figure}[tbp]
\caption{Phase diagrams in the planes $\varepsilon$ vs.
$\alpha_i$. a) first layer, b) second layer. Region I corresponds
to the dilute (vapor-like) phase, region III to the dense
(liquid-like) phase, while II corresponds to the region of phase
coexistence.}
\end{figure}

\begin{figure}[tbp]
\caption{Typical form of the isotherm. We depict the total
coverage $\theta (= c_1 + c_2)$  vs. $\varphi$, the chemical
potential.}
\end{figure}

\begin{figure}[tbp]
\caption{Here we depict the relation between the step size and
the potential parameter $\varepsilon $: $\Delta \varphi$ vs.
$\varepsilon$, yielding a linear dependence in logarithmic scale.
The result here corresponds to the 1st step. }
\end{figure}

\begin{figure}[tbp]
\caption{Typical form of the density profile. Here we have used
$\varepsilon = 5.6$, $\alpha _1 = 0.08$  and $\alpha _2 = 0.096$}
\end{figure}

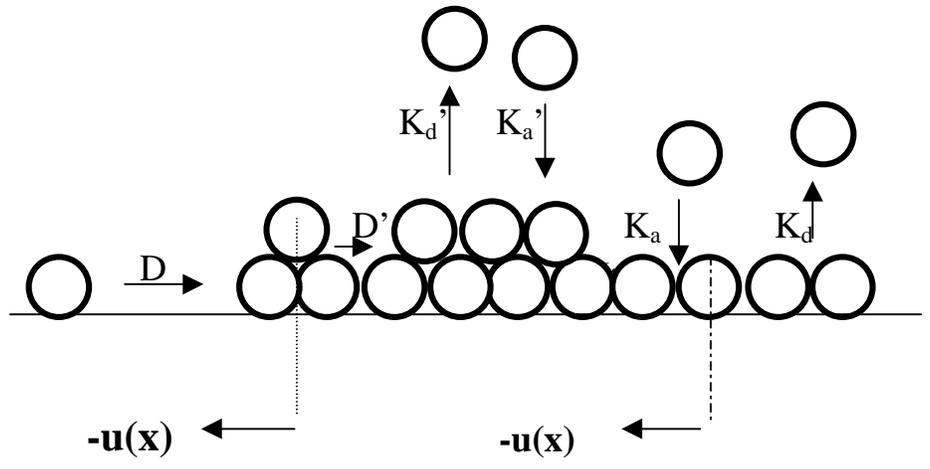
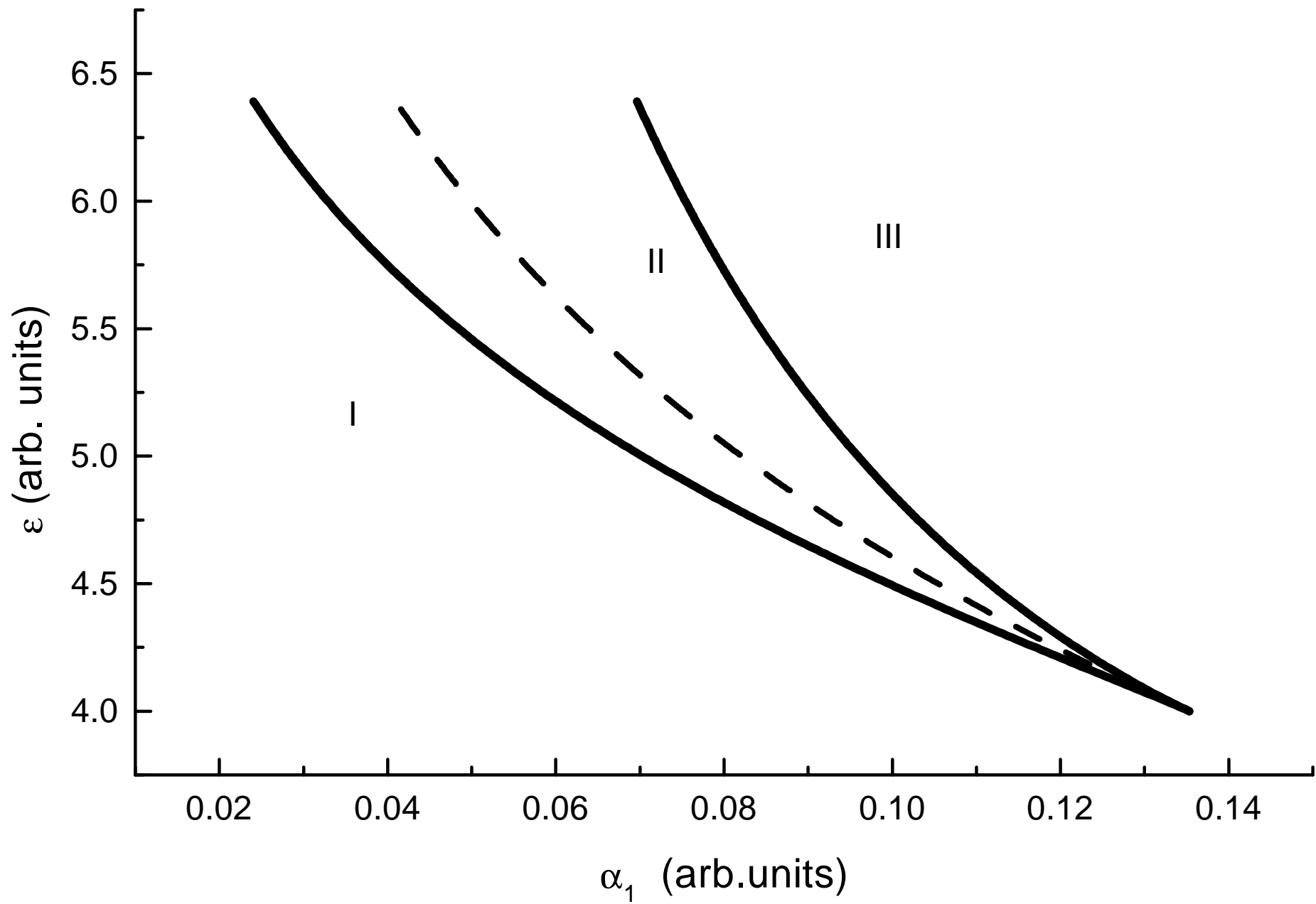
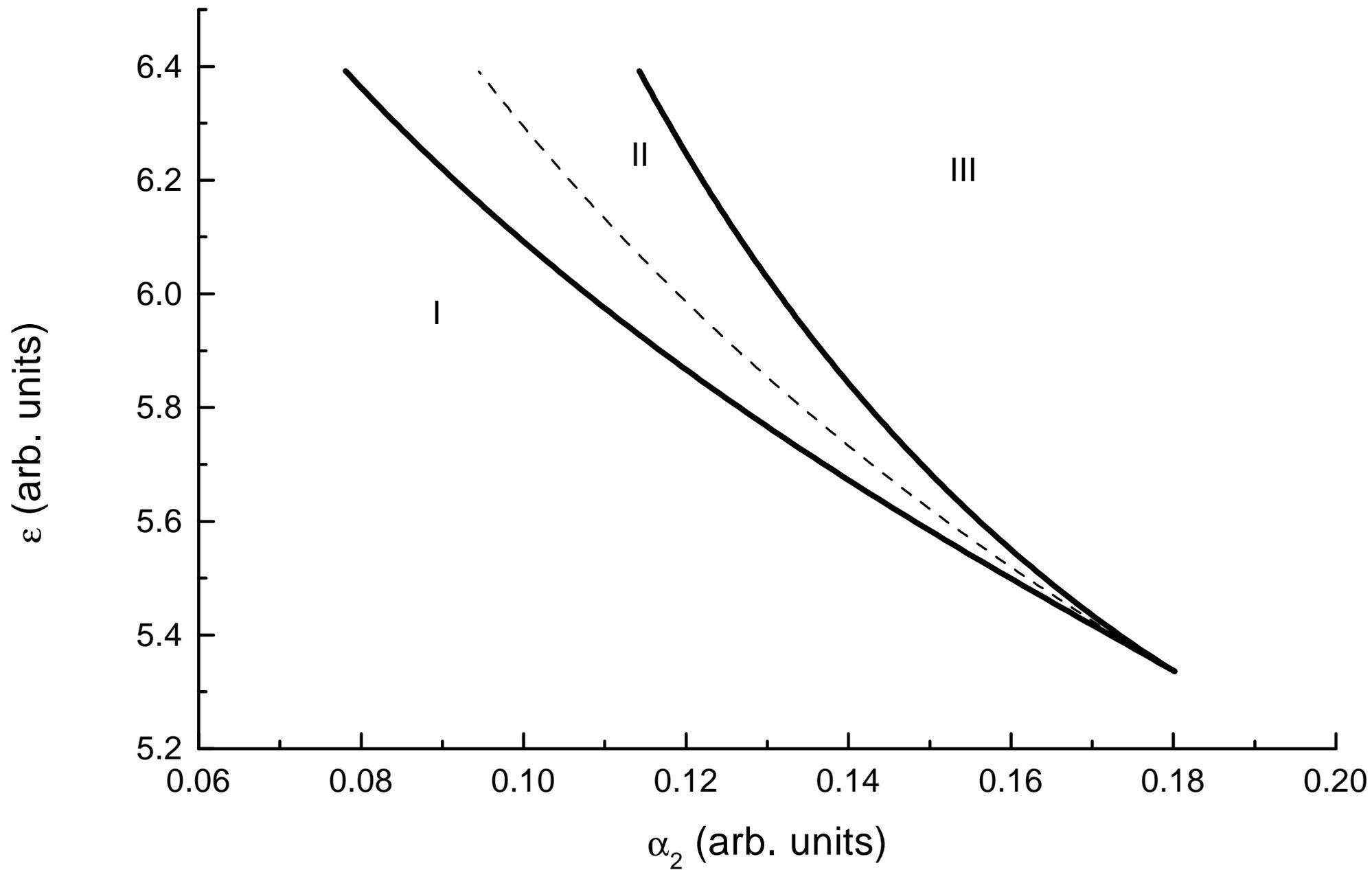
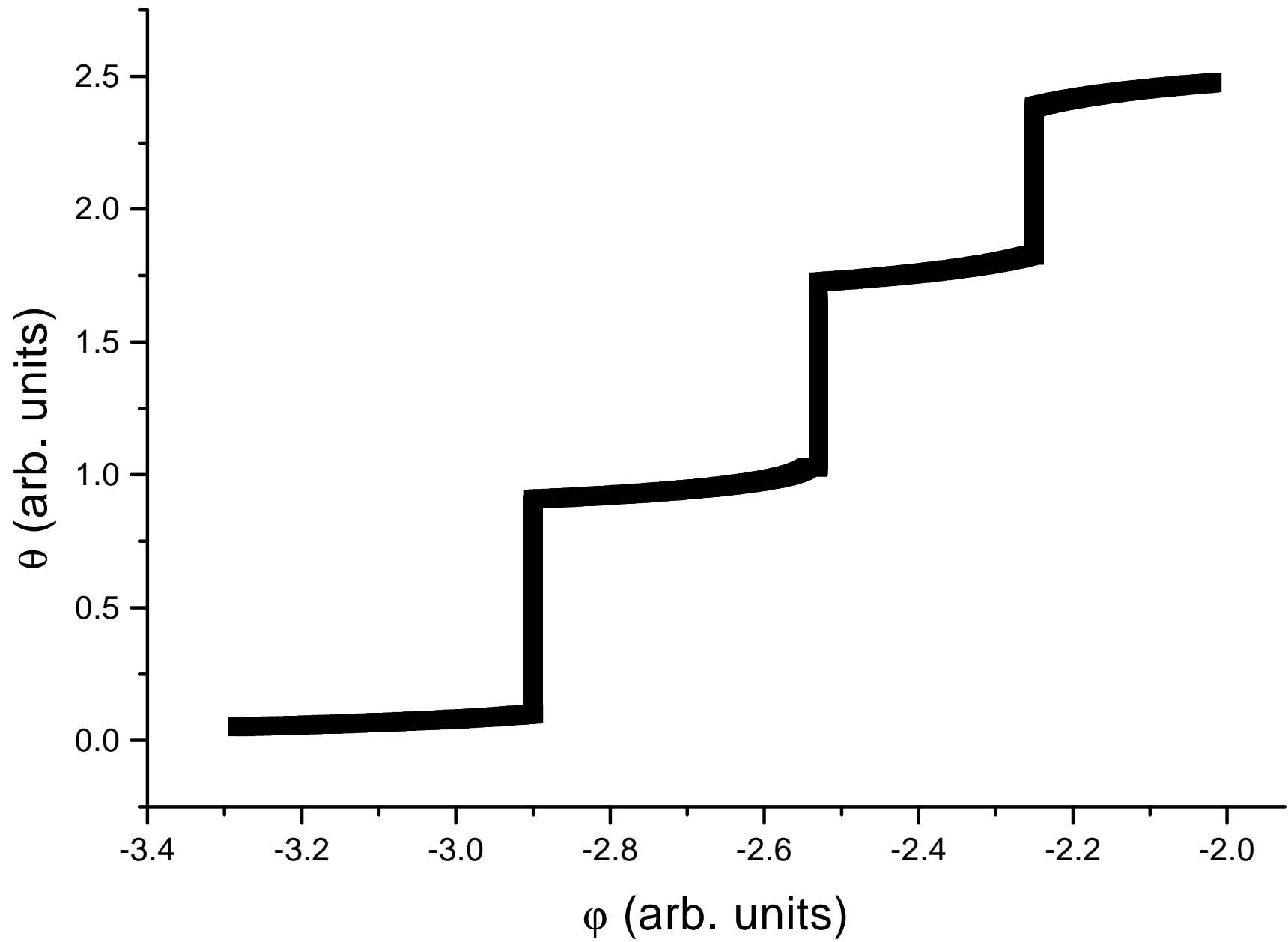
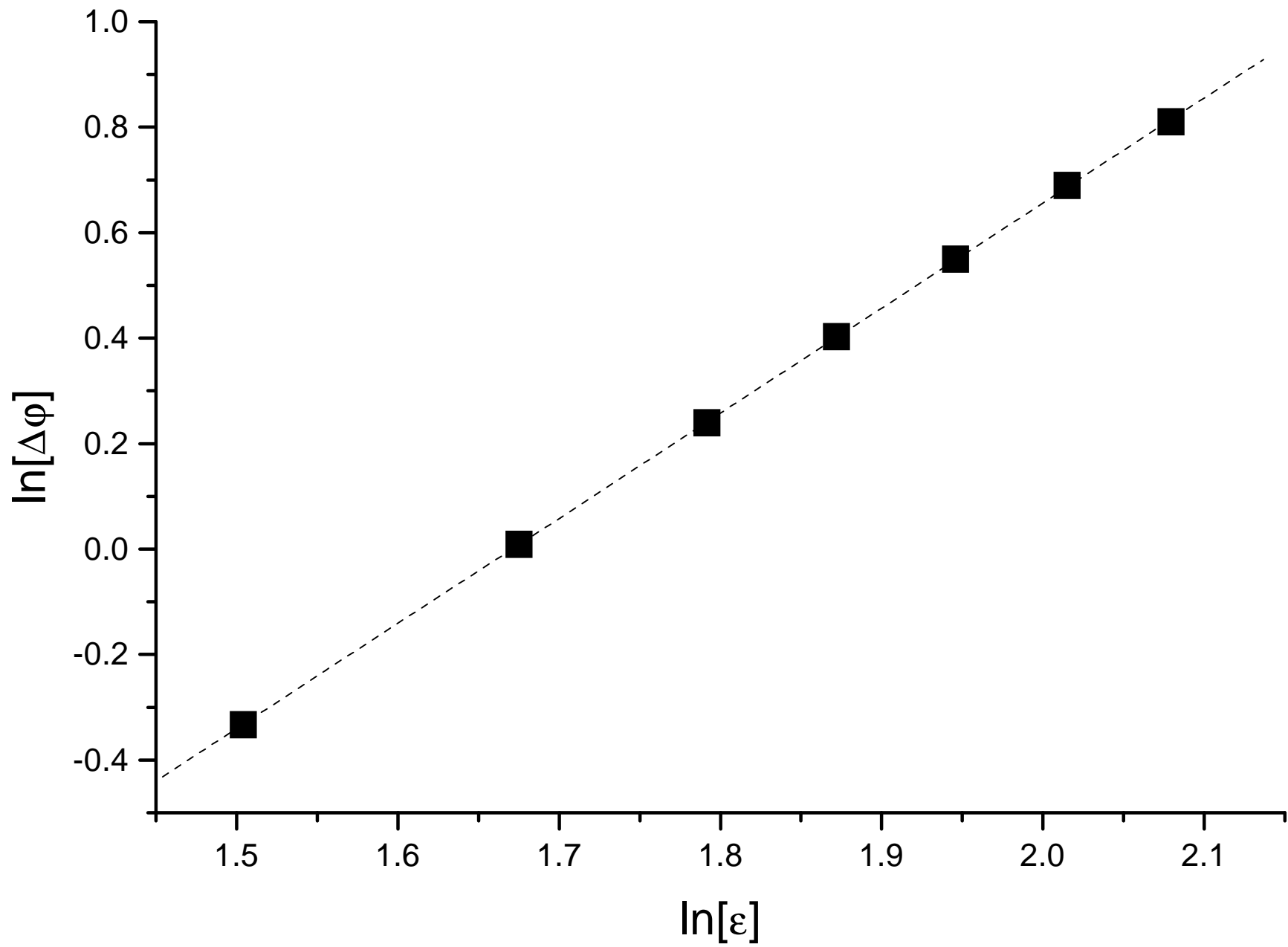
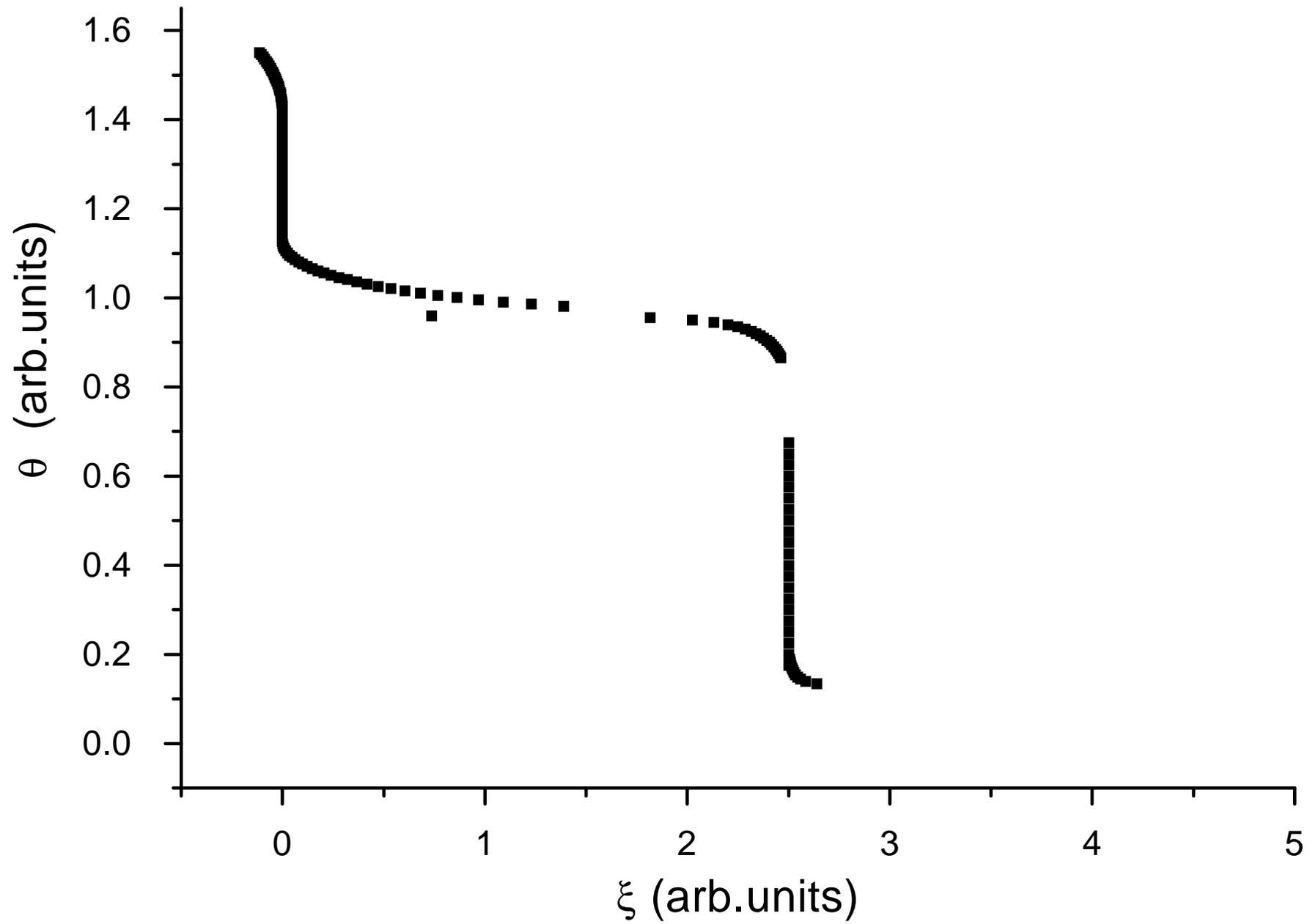
\end{document}